\documentclass[aps,prl,twocolumn,amsfonts,amssymb,amsmath,showpacs,
floatfix,nofootinbib,groupedaddress,superscriptaddress,citesort]{revtex4}
\usepackage{mathrsfs}
\usepackage{amsfonts}
\usepackage{amstext}
\usepackage{amsmath}
\usepackage{amssymb}
\usepackage{bm}
\usepackage{bbm}
\usepackage[dvips]{graphicx}
\def\qed{\leavevmode\unskip\penalty9999 \hbox{}\nobreak\hfill
     \quad\hbox{\leavevmode  \hbox to.77778em{%
              \hfil\vrule   \vbox to.675em%
               {\hrule width.6em\vfil\hrule}\vrule\hfil}}
     \par\vskip3pt}

\def\ra{\rangle}
\def\la{\langle}

\begin{document}

\title{Unified view of monogamy relations for different entanglement measures\\}

\author{Yu Guo}
\email{guoyu3@aliyun.com}
\affiliation{Institute of Quantum Information Science, Shanxi Datong University, Datong, Shanxi 037009, China}
\affiliation{School of Mathematics and Computer Science, Shanxi Datong University, Datong, Shanxi 037009, China}%


\begin{abstract}

A particularly interesting feature of nonrelativistic
quantum mechanics is the monogamy laws of entanglement.
Although the monogamy relation has been explored extensively in the last decade,
it is still not clear to what extent a given entanglement measure is monogamous.
We give here a conjecture on the amount of entanglement contained in the reduced states
by observing all the known related results at first.
Consequently, we propose the monogamy power of an entanglement measure and
the polygamy power for its dual quantity, the assisted entanglement,
and show that both the monogamy power and the
polygamy power exist in any multipartite systems with any dimension,
from which we formalize exactly for the first time when an entanglement measure and an
assisted entanglement obey the monogamy relation
and the polygamy relation respectively
in a unified way.
In addition, we show that any entanglement measure violates the polygamy relation,
which is misstated in some recent papers.
Only the existence of monogamy power is conditioned on the conjecture,
all other results are strictly proved.

\end{abstract}

\maketitle

Monogamy law of entanglement as one of the most striking features in quantum
world has been explored ever since the distribution of three qubit state entanglement
discovered by Coffman, Kundu, and Wootters (CKW) \cite{Horodecki2009,Coffman,Koashi,
Gour2005pra,Osborne,Adesso2006njp,Adesso2006pra,Gour,Ouyongcheng,
Ouyongcheng2007pra2,Hiroshima2007prl,Adesso,Kim,Yuchangshui2008pra,
Adesso2008pra,Kim2009,Buscemi,Kim2009pra,Lizongguo,
Kim2010jpa,Renxijun,Cornelio,Streltsov,Kim2012pra,Braga,Cornelio2013pra,
Liusiyuan,Reid,Kim2014epjd,Zhuxuena2014pra,Bai,Regula2014prl,Salini,
Choi,Luo,Hehuan,Eltschka,Kumar2015,Zhuxuena2015pra,Audenaert,Kumar,
Lancien,Lami,Song,Regula,Luo2016pra,Kumar2016pra,Jia,Jung,Chengshuming2016pra,Kim2016pra,
Zhuxuena2017qip,Chengshuming,Allen}.
It requires restricted shareability of entanglement, i.e., the more two
particles are entangled the less this pair have entanglement with the rest,
which sets quantum correlation apart from classical correlations.
This feature has found potential applications in
quantum information tasks and and other areas of physics, such as
quantum key distribution \cite{Terhal,Pawlowski,Gisin},
classifying quantum states \cite{Dur,Giorgi,Prabhu},
condensed-matter physics \cite{Ma,Brandao,Garcia},
frustrated spin systems \cite{Rao}, statistical physics \cite{Bennett}, and even
black-hole physics \cite{Susskind,Lloyd}.

Monogamy relation is displayed as an inequality quantitatively.
We recall the monogamy inequality in a large scale, namely, not only
for the entanglement, but also for any other bipartite quantum
correlation measures.
Let $E$ be a quantity that quantifying some bipartite correlation.
We denote the state of a multipartite system with state space
$H_a\otimes H_{b_1}\otimes H_{b_2}\otimes\cdots\otimes H_{b_n}:=H_{a\textbf{b}}$
 by $\rho_{ab_1b_2\cdots b_n}:=\rho_{a\emph{\bf{b}}}$, $\dim H_{a\textbf{b}}<\infty$.
 The set of all states acting on
 $H_{a\textbf{b}}$
 is denoted by $\mathcal{S}_{a\textbf{b}}$.
$E$ is said to be \emph{monogamous} if
\begin{eqnarray}
E(\rho_{a|\textbf{b}})\geq\sum_{i=1}^nE(\rho_{a|b_i})
\label{monogamydefinition}
\end{eqnarray}
holds for any $\rho_{a\textbf{b}}\in\mathcal{S}_{a\textbf{b}}$,
where the vertical bar denotes the bipartite split across which it is computed.
Or else, it is \emph{polygamous}.

In general, entanglement measure $E$ always violates
Eq.~(\ref{monogamydefinition}) while $E^\alpha$ satisfies the monogamy relation for some $\alpha>1$
\cite{Kumar,Zhuxuena2014pra,Zhuxuena2015pra,Coffman,Osborne,Kim2009,Luo,Bai,Ouyongcheng2007pra2,Choi}.
Recently, Kumar showed in Ref.~\cite{Kumar} that
monogamy is preserved for raising the power and polygamy is maintained for lowering the power.
Salini \emph{et al} found that the monotonically increasing function of any
monotonic quantum correlation can make all multiparty
states monogamous \cite{Salini}.
This raises the following issues:
For an arbitrarily given entanglement measure $E$,
can we find a simplest function $f$ such that $f(E)$ is monogamous for all states?
If so, how can we determine the sharp one that leaves it monogamous?
In addition, the assisted entanglement, which is a dual amount of entanglement measure, has shown to be polygamous
with respect to some entanglement measures,
such as entanglement of assistance \cite{Kim2012pra,DiVincenzo},
tangle of assistance \cite{Gour2005pra,Gour}, concurrence of assistance \cite{Gour},
Tsallis entropy of assistance \cite{Kim2016pra}
and convex roof extended negativity of assistance \cite{Kim2009}.
Then, what about the issues above for polygamy relation of assisted entanglement?

The main aim of this Letter is to address these issues.
We begin by observing all the known results about monogamy relation of entanglement, from which
we conjecture that: if the amount of entanglement in a multipartite state
is equal to one of its reduced states, then all the other reduced states are separable.
Based on this conjecture,
we concentrate on the simplest monotonically increasing function $f(x)=x^\alpha$
and then define the infimum $\alpha$ such that $E^\alpha$ is still monogamous
as \emph{monogamy power} of the given measure $E$.
We also discuss the monogamy of the convex roof extended $E^\alpha$,
denoted by $\tilde{E}^\alpha$, and the relation between $E^\alpha$ and $\tilde{E}^\alpha$.
The second part touches on the dual relation of monogamy,
i.e., the polygamy of assisted entanglement $E_a^\beta$, with the same approach
but dual to that of the monogamy for $E^\alpha$.

For the one-way distillable entanglement $E_d$ \cite{Devetak} and the squashed entanglement $E_{sq}$ \cite{Christandl},
the monogamy relation holds for any tripartite systems \cite{Koashi}.
This yields that
$E(\rho_{a|bc})=E(\rho_{a|b})$, then $E(\rho_{a|c})=0$,
for $E=E_d$ or $E=E_{sq}$.
He and Vidal proved in Ref.~\cite{Hehuan}
that, if $N(|\psi\ra_{a|bc})=N(\rho_{a|b})$,
then $N(\rho_{ac})=0$, where $N$ is the negativity \cite{Vidal2002pra}, $|\psi\ra_{abc}$ is any tripartite
pure state.
In addition, all the results in Refs.~\cite{Zhuxuena2014pra,
Osborne,Ouyongcheng,Renxijun,Kim,Ouyongcheng2007pra2,Luo,Kim2009,
Choi,Kumar,Bai,Kim2016pra,Luo2016pra,Kim2010jpa,Cornelio,Song}(see Table~\ref{tab:table1})
display the same characteristic. We thus present the following conjecture.

{\it Conjecture.}
Let $E$ be an entanglement measure. If
$E(\rho_{a|\textbf{b}})=E(\rho_{a|b_{i_o}})$ for some $i_0$,
then $E(\rho_{a|b_i})=0$ for any $i\neq i_0$.

Conditioned on this conjecture, we give the first result of this Letter. It confirms
that for any given entanglement measure $E$ (no matter it is monogamous or not)
there exists $\alpha>0$ such that the power function $E^\alpha$ is always monogamous, and
the existence is irrespective of both the number of subsystems and
the dimensions.

{\it Theorem 1.} For any well-defined bipartite entanglement measure $E$,
there exist $\alpha>0$ such that
\begin{eqnarray}
E^\alpha(\rho_{a|\bf{b}})\geq\sum_{i=1}^nE^\alpha(\rho_{a|b_i})
\label{monogamy-inequ-1}
\end{eqnarray}
holds for any $\rho_{ab}\in\mathcal{S}_{a\textbf{b}}$.

{\it Proof.} Since $E$ is an entanglement monotone,
it is nonincreasing under partial traces,
i.e., $E(\rho_{a|\textbf{b}})\geq E(\rho_{a|b_i})$ for any $1\leq i\leq n$.
Let $E(\rho_{a|\textbf{b}})=x$ and $E(\rho_{a|b_i})=y_i$, $1\leq i\leq n$.
We assume with no loss of generality that $n=2$.
By Conjecture, $x>y_i$, $i=1$, 2, or $x=y_1$ and $y_2=0$, or $x=y_2$ and $y_1=0$,
this guarantees that there exists $\alpha>0$ such that $x^\alpha\geq y_1^\alpha+y_2^\alpha$.
\hfill$\square$

That is, any entanglement $E$ can deduce an quantity $E^\alpha$ which satisfies the
monogamy relation even though $E$ is not monogamous itself.
For example, concurrence $C$ is not monogamous,
but $C^2$ is \cite{Osborne,Kim2009} (also see in Table~\ref{tab:table1}).
We now expect to ascertain the domain of the power $\alpha$ which admits the monogamy relation.
Let $\rho_{a\textbf{b}}\in\mathcal{S}_{a\textbf{b}}$
and $E$ be bipartite normalized entanglement measure (i.e., $0\leq E\leq 1$).
It is shown in Ref.~\cite{Kumar} that $E^r(\rho_{a|\textbf{b}})\geq\sum_iE^r(\rho_{a|b_i})$
implies $E^\alpha(\rho_{a|\textbf{b}})\geq\sum_iE^\alpha(\rho_{a|b_i})$
for $\alpha\geq r\geq 1$.
In fact, it is still valid for unnormalized entanglement measure
since $(x+y)^\alpha\geq x^\alpha+y^\alpha$ for $\alpha\geq1$ and $x,y\geq 0$.
That is, for any given entanglement measure $E$, the corresponding minimal power index $\alpha$
reflects its monogamy in nature.
This motivates the following definition.

{\it Definition 1.}
Let $E$ be a bipartite entanglement measure. For a given multipartite system described by
$H_{a\textbf{b}}$,
we define the \emph{monogamy power} of $E$ by
\begin{eqnarray}
\alpha(E):&=&\inf\{\alpha: E^\alpha(\rho_{a|\textbf{b}})\geq\sum_{i=1}^nE^\alpha(\rho_{a|b_i})\nonumber\\
 &&~~~~~~\mbox{for all}\ \rho_{a\textbf{b}}\in\mathcal{S}_{a\textbf{b}}\},
\label{monogamy-index}
\end{eqnarray}

That is, $\alpha(E)$ is the infimum exponent for $E^{\alpha(E)}$ to be monogamous for the given measure $E$.
In fact, $\alpha(E)$ is also dependent on the size of the systems (see in Table~\ref{tab:table1}).
All the known research results can imply that
$\alpha(E)$ is hard to compute due to the complexity of monogamy relation especially for higher dimensions
\cite{Hehuan,Lancien,Allen,Eltschka,Audenaert,Chengshuming}.
From Table~\ref{tab:table1}, we may conjecture that $\alpha(E)$ will largen with the increasing of the dimension of the subsystems.

Reference~\cite{Kumar} also showed that $Q^r(\rho_{a|\textbf{b}})\leq \sum_iQ^r(\rho_{a|b_i})$
implies $Q^\alpha(\rho_{a|\textbf{b}})\leq\sum_iQ^\alpha(\rho_{a|b_i})$
for $\alpha\leq r$, where $Q$ is any given normalized bipartite correlation measure.
However, it is conditioned on the given state, it is not true for
all states if $Q$ is an entanglement measure.
For example, we consider the following pure state
\begin{eqnarray}
|\psi\ra_{a\textbf{b}}=\sum_{j=0}^{k-1}\lambda_j|e_j^{(0)}\rangle|e_j^{(1)}\rangle\cdots|e_j^{(n)}\rangle,
\label{ghz-class}
\end{eqnarray}
 where $\{|e_j^{(0)}\rangle\}$ is an orthonormal
set in $H_a$, and $\{|e_j^{(i)}\rangle\}$ is an orthonormal set of $H_{b_i}$,
$\sum_j\lambda_j^2=1$, $\lambda_j >0$,
$k\leq\min\{\dim H_a,\dim H_{b_1},\dots,\dim H_{b_n}\}$,
$i=1$, 2, $\dots$, $n$, $n\geq 3$
(these states admit the multipartite Schmidt decomposition \cite{Guoyu2015jpa,Guoyu2015qip}).
We always have $E(|\psi\ra_{a|\textbf{b}})>0$ while $E(\rho_{a|b_i})=0$ for all $1\leq i\leq n$.
On the other hand, for the three qubit state
\begin{eqnarray}
|\phi\ra_{a\textbf{b}}=\frac{1}{\sqrt{5}}(\sqrt{2}|100\ra+\sqrt{2}|110\ra+|111\ra),
\label{example2}
\end{eqnarray}
we can easily calculate that
$C(|\phi\ra_{a|\textbf{b}})\approx0.9798$, $C(\rho_{a|b_1})\approx0.5656$ and $\rho_{a|b_2}$ is separable.
This give rise to the following theorem.

{\it Theorem 2.} For any well-defined bipartite entanglement measure $E$,
there does not exist $\beta$ such that
\begin{eqnarray}
E^\beta(\rho_{a|\textbf{b}})\leq\sum_{i=1}^nE^\beta(\rho_{a|b_i})
\label{monogamy-inequ-2}
\end{eqnarray}
hold for any $\rho_{a\textbf{b}}\in\mathcal{S}_{a\textbf{b}}$.

For $2\otimes 2\otimes 2^{m}$ systems, $m\geq 1$, it is shown in Ref.~\cite{Luo}
that $N^\beta(|\psi\ra_{a|\textbf{b}})\leq N^\beta(\rho_{a|b_1})+N^\beta(\rho_{a|b_2})$
and $E^\beta(\rho_{a|\textbf{b}})\leq E^\beta(\rho_{a|b_1})+E^\beta(\rho_{a|b_2})$
for $E=N_{cr}$  or $E=E_f$ whenever $\beta\leq0$.
However, these statements are not true from the states in Eq.~(\ref{ghz-class}).

\begin{table}
\caption{\label{tab:table1}The comparison of the monogamy power of several entanglement.
We denote the one-way distillable entanglement, concurrence, negativity, convex roof extended negativity, entanglement of formation,
tangle, squashed entanglement, Tsallis-$q$ entanglement and R\'{e}nyi-$\alpha$ entanglement by $E_d$,
$C$, $N$,$N_{cr}$, $E_f$, $\tau$, $E_{sq}$, $T_q$ and $R_\alpha$, respectively.}
\begin{ruledtabular}
\begin{tabular}{cccc}
 $E$& $\alpha(E)$ & system & reference \\ \colrule
 $E_d$ & $\leq1$& any system & \cite{Koashi}\\
$C$ & $\leq\sqrt{2}$& $2^{\otimes n}$ & \cite{Zhuxuena2014pra}\\
                 &   $\leq 2$    & $2\otimes 2\otimes2^m$ &\cite{Osborne}\\
                 &   $\leq 2$    & $2^{\otimes n}$ &\cite{Osborne}\\
                 &   $>2$        & $3^{\otimes 3}$ & \cite{Ouyongcheng}\\
                 &   $\leq 2$    & $2\otimes 2\otimes4$ &\cite{Renxijun} \\
                 &   $>3$    & $3\otimes 2\otimes2$ &\cite{Kim} \\
  $N$& $\leq 2$ & $2^{\otimes n}$ &\cite{Ouyongcheng2007pra2,Luo}\footnotemark[1]\\
          & $\leq 2$ & $2\otimes 2\otimes2^m$ &\cite{Luo}\footnotemark[1]\\
          & $\leq 2$ & $d\otimes d\otimes d$,$d=2,3,4$ &\cite{Hehuan}\\
 $N_{cr}$ & $\leq 2$ & $2^{\otimes n}$ & \cite{Kim2009,Luo,Choi}\\
$E_f$& $\leq 2$ & $2^{\otimes n}$ & \cite{Kumar,Bai}\\
& $\leq \sqrt{2}$ & $2^{\otimes n}$ & \cite{Zhuxuena2014pra}\\
 $\tau$ &   $\leq 1$    & $2\otimes 2\otimes4$ &\cite{Renxijun}\\
$E_{sq}$& $\leq 1$& any system &\cite{Koashi}\\
$T_{q}$, $2\leq q\leq 3$& $\leq 1$& $2^{\otimes n}$ &\cite{Kim2016pra}\\
$T_{q}$& $\leq 2$& $2\otimes 2\otimes2^m$ &\cite{Luo2016pra}\footnotemark[2]\\
$R_{\alpha}$, $\alpha\geq2$& $\leq 1$& $2^{\otimes n}$ &\cite{Kim2010jpa,Cornelio}\\
$R_{\alpha}$, $\alpha\geq\frac{\sqrt{7}-1}{2}$& $\leq 2$& $2^{\otimes n}$ &\cite{Song}\footnotemark[3]
\end{tabular}
\end{ruledtabular}
\footnotetext[1]{For pure states.}
\footnotetext[2]{For mixed states, and $q\in[\frac{5-\sqrt{13}}{2},2]\cup[3,\frac{5+\sqrt{13}}{2}]$.}
\footnotetext[3]{For mixed states.}
\end{table}

Kumar proved in Ref.~\cite{Kumar} that, if $E$ is a normalized
convex roof extended bipartite entanglement measure and monogamous for pure state,
then $E$ and $E^2$ are monogamous for mixed states.
(Here, we call $E$ is convex roof extended if
$E(\rho_{ab}):=\inf_{\{p_i,\psi_i\}}\sum_ip_iE(|\psi_i\ra)$ for any mixed state $\rho_{ab}\in\mathcal{S}_{ab}$
provided that $E$ is originally defined for pure states,
where the infimum
is taken over all possible ensembles $\{p_i,|\psi_i\rangle\}$ of $\rho_{ab}$.)
We can derive more indeed.

{\it Proposition 1.} Let $E$ be a
convex roof extended bipartite entanglement measure. If it is monogamous for pure state,
then $E^\alpha$ is monogamous for both pure and mixed states for any $\alpha\geq1$.

{\it Proof.} By Theorem 4 in Ref.~\cite{Kumar},
the monogamy of $E$ for pure state implies $E$ is monogamous for mixed state (note that
it is also true without the normalized condition).
Then the monogamy of $E^\alpha$ is straightforward since $\alpha\geq1$.
\hfill$\square$

For convenience and completeness, we list below the hierarchical monogamy relations which have been proposed in
Refs.~\cite{Horodecki2009,Osborne,Kumar2015,Kumar}.
Based on this hierarchy, we only need to consider monogamy relations for the tripartite case in nature.

{\it Proposition 2. \cite{Horodecki2009,Osborne,Kumar2015,Kumar}}
Let $E$ be a convex roof extended bipartite entanglement measure.
If $E^{\alpha}$ is monogamous for any tripartite system, then
$E^\gamma$ is monogamous for any $n$-partite system $n\geq3$ and any $\gamma\geq\alpha$.
If $E$ is monogamous for all tripartite pure
states in $d\otimes d\otimes d'$ system, $d'=d^m$, $2\leq m\leq n-2$,
then it is monogamous for all $d^{\otimes n}$ states.

A natural question is whether the convex roof extended measure deduced from $E^\alpha$ is
monogamous too?
For simplicity,
if $E$ is a bipartite entanglement measure,
we define
\begin{eqnarray}
\tilde{E}^\alpha(|\psi\ra_{ab}):=E^\alpha(|\psi\ra_{ab})
\end{eqnarray}
for any pure state $|\psi\ra_{ab}\in H_a\otimes H_b$
and
\begin{eqnarray}
\tilde{E}^\alpha(\rho_{ab}):=\inf_{\{p_i,\psi_i\}}\sum_ip_i\tilde{E}^\alpha(|\psi_i\ra),\ \rho_{ab}\in\mathcal{S}_{ab},
\label{crm}
\end{eqnarray}
where the infimum
is taken over all possible ensembles $\{p_i,|\psi_i\rangle\}$ of $\rho_{ab}$.
For example, $\tilde{C}^2=\tau$ and $\tilde{N}=N_{cr}$.

{\it Proposition 3.} If $\tilde{E}^\alpha$ is monogamous for pure states, then
it is also monogamous for mixed states.

{\it Proof.} We only need to check it for the tripartite case.
Let $\rho_{a\textbf{b}}$ be any given mixed state in $\mathcal{S}_{a\textbf{b}}$.
For any $\epsilon>0$, there exits an ensemble
$\{p_i,|\psi_i\ra\}$ such that
$\tilde{E}^\alpha(\rho_{a|\textbf{b}})\geq\sum_ip_i\tilde{E}^\alpha(\rho^{(i)}_{a|\textbf{b}})-\epsilon$,
where $\rho^{(i)}_{ab}=|\psi_i\ra\la\psi_i|$.
It follows that
$\tilde{E}^\alpha(\rho_{a|\textbf{b}})\geq\sum_ip_i\tilde{E}^\alpha(\rho^{(i)}_{a|\textbf{b}})-\epsilon
\geq\sum_ip_i[\tilde{E}^\alpha(\rho^{(i)}_{a|b_1})+\tilde{E}^\alpha(\rho^{(i)}_{a|b_2})]-\epsilon
\geq\tilde{E}^\alpha(\rho_{a|b_1})+\tilde{E}^\alpha(\rho_{a|b_2})-\epsilon$.
Therefore, $\tilde{E}^\alpha(\rho_{a|\textbf{b}})\geq\tilde{E}^\alpha(\rho_{a|b_1})+\tilde{E}^\alpha(\rho_{a|b_2})$
since $\epsilon$ is arbitrary.
\hfill$\square$

Note that the monogamy of $E^\alpha$ for pure states does not imply $E^\alpha$ is monogamous for
mixed states if $E$ is not a convex roof extended measure in general.
For the case of $\alpha=2$, the monogamy of $\tilde{E}^2$
is stronger than $E^2$.

{\it Proposition 4.} Let $E$ be a convex roof extended bipartite entanglement measure.
If $\tilde{E}^2$  is monogamous, then
so is $E^2$.

{\it Proof.} We only need to prove $E^2\leq \tilde{E}^2$.
Let $\rho_{ab}$ be any state in $\mathcal{S}_{ab}$.
For any $\epsilon>0$, there exists
$\{p_i,|\psi_i\rangle\}$ such that
$\rho_{ab}=\sum\limits_ip_i|\psi_i\rangle\langle\psi_i|$ and
$\tilde{E}^2(\rho_{ab})\geq\sum\limits_ip_iE^2(|\psi_i\rangle)-\epsilon$.
 Then we have
$E^2(\rho_{ab})\leq[\sum\limits_ip_iE(|\psi_i\rangle)]^2
=[\sum\limits_i\sqrt{p_i}\sqrt{p_i}E(|\psi_i\rangle)]^2
\leq[\sum\limits_ip_i][\sum\limits_ip_iE^2(|\psi_i\rangle)]
\leq\tilde{E}^2(\rho_{ab})+\epsilon$,
which establishes the inequality $E^2\leq\tilde{E}^2$ since
$\varepsilon$ is arbitrary.\hfill$\square$

For example, the $2\otimes 2$ pure state $|\psi\ra$, $C(|\psi\ra)=N(|\psi\ra)$,
so we have
\begin{eqnarray}
\tilde{N}^2(\rho_{a|bc})\geq\tilde{N}^2(\rho_{a|b})+\tilde{N}^2(\rho_{a|c})
\end{eqnarray}
for any three-qubit state $\rho_{abc}$ (note that $\tilde{N}^2=\tilde{N}_{cr}^2$).
Therefore
\begin{eqnarray}
N_{cr}^2(\rho_{a|bc})\geq N_{cr}^2(\rho_{a|b})+N_{cr}^2(\rho_{a|c})
\end{eqnarray}
as discussed in Ref.~\cite{Kim2009,Choi}, which is more clear
from Proposition 4.
However, the converse of Proposition 4 may be not true.
It is worth noticing that $\tilde{E}^\alpha$ is not a entanglement measure
since the concavity of a function $f(\rho_a):=E(|\psi\ra_{ab})$ can not guarantee $f^\alpha$
is concave in general under the scenario in Ref.~\cite{Vidal}.

The rest of this Letter will devote to discuss the polygamy of
the assisted entanglement, i.e.,
the dual concept of monogamy of entanglement.
We present the discussion following the frame of monogamy part above.
Recall that, the first assisted entanglement is the \emph{entanglement of assistance} which is dual quantity of
the entanglement formation \cite{DiVincenzo}
\begin{eqnarray}
{E_f}_a(\rho_{ab})=\sup_{\{p_i,|\psi_i\ra\}}\sum_ip_iE_f(|\psi_i\ra), \ \rho_{ab}\in\mathcal{S}_{ab},
\end{eqnarray}
where the supremum is taken over all ensembles $\{p_i,|\psi_i\ra\}$ of $\rho_{ab}$.
In general, for any given bipartite entanglement measure $E$,
the corresponding assisted entanglement is defined by
\begin{eqnarray}
E_a(\rho_{ab})=\sup_{\{p_i,|\psi_i\ra\}}\sum_ip_iE(|\psi_i\ra),\ \rho_{ab}\in\mathcal{S}_{ab}.
\end{eqnarray}
We firstly show that some power of $E_a$ can be polygamous,
which reveals the dual property of the entanglement measure $E$.

{\it Theorem 3.} For any bipartite entanglement measure $E$,
there exist $\beta>0$ such that
\begin{eqnarray}
E_a^\beta(\rho_{a|\textbf{b}})\leq\sum_{i=1}^nE_a^\beta(\rho_{a|b_i})
\label{polygamy-inequ-1}
\end{eqnarray}
holds for any $\rho_{a\textbf{b}}\in\mathcal{S}_{a\textbf{b}}$.

{\it Proof.} We assume with no loss of generality that $n=2$.
If $\rho_{a\textbf{b}}=|\psi\ra_{ab}\la\psi|$ and it is $a|\textbf{b}$ separable,
it is clear for any $\beta>0$.
It remains to show that
$E_a(\rho_{ab_i})>0$ for any mixed state $\rho_{ab_i}$, $i=1$, 2,
since there always exists $\beta>0$ such that $x^\beta\leq y_1^\beta+y_2^\beta$
for and $x\geq0$ and $y_i>0$.
Let $\rho_{ab_i}=\sum_j\lambda_j|\psi_j^{(i)}\ra\la\psi_j^{(i)}|$ be its spectrum decomposition.
If there exists $|\psi_{j_0}^{(i)}\ra$ is entangled for some $j_0$, then $E_a(\rho_{ab_i})>0$.
Or else, we assume that all $|\psi_{j}^{(i)}\ra$s are separable.
Without losing of generality, let
$\rho_{ab_i}=\sum_{j=1}^2\lambda_j|j\ra|j\ra\la j|\la j|$,
then $\rho_{ab_i}$ can be rewritten as
$\rho_{ab_i}=|\psi_1\ra\la\psi_1|+|\psi_2\ra\la\psi_2|$,
where $|\psi_1\ra=cd|1\ra|1\ra+e|2\ra|2\ra$ and
$|\psi_2\ra=d|1\ra|1\ra+ce|2\ra|2\ra$ (here, $|\psi_{1,2}\ra$ is unnormalized) with
$(1+c)d^2=\lambda_1$ and $(1+c)e^2=\lambda_2$.
This reveals $E_a(\rho_{ab_i})>0$.
\hfill$\square$

Conversely, we conjecture that
there is no monogamy relation for the assisted entanglement
since $E_a$ is not a well-defined entanglement measure in general \cite{Gour2006pra}.
On the contrary to monogamy preserving for rasing power,
the polygamy of $E_a^\beta$ can also be preserved when we lower the power
from Theorem 2 in Ref.~\cite{Kumar}:
Let $\rho_{a\textbf{b}}\in\mathcal{S}_{a\textbf{b}}$ and $E_a$
be a bipartite assisted entanglement.
Then $E_a^\beta(\rho_{a|\textbf{b}})\leq \sum_iE_a^\beta(\rho_{a|b_i})$
implies $E_a^\gamma(\rho_{a|\textbf{b}})\leq\sum_iE_a^\gamma(\rho_{a|b_i})$
for $0\leq\gamma\leq \beta$.
(Note that, in Ref. \cite{Kumar}, the bipartite correlation measure $Q$
is assumed to be normalized. However this condition is not necessary,
which can also be checked easily following
the argument therein.)
The argument above thus guarantee that the following concept is well-defined.

{\it Definition 2.}
Let $E$ be a bipartite entanglement measure. For a given multipartite system described by
$H_a\otimes H_{b_1}\otimes H_{b_2}\otimes\cdots\otimes H_{b_n}$,
we call
\begin{eqnarray}
\beta(E_a):&=&\sup\{\beta: E_a^\beta(\rho_{a|\textbf{b}})\leq\sum_{i=1}^nE_a^\beta(\rho_{a|b_i})\nonumber\\
 &&~~~~~~\mbox{for all}\ \rho_{a\textbf{b}}\in\mathcal{S}_{a\textbf{b}}\}
\label{polygamy-index}
\end{eqnarray}
the \emph{polygamy power} of the assisted entanglement $E_a$, i.e.,
$\beta(E_a)$ is the supremum for $E^{\beta(E)}$ to be polygamous for the given entanglement measure $E$.

Together with Definition 1, the pair $(\alpha(E),\beta(E_a))$
reflects the shareability of entanglement for $E$ completely.
This pair of power indexes advances our understanding of multipartite entanglement
although theses quantities are difficult to calculate.
In addition, some related problems are straightforward:
Is there a close connection between $\alpha(E)$ and $\beta(E_a)$?
Is $\beta(E_a)$ dependent on the size of state space?
These issues deserve further study (some known examples are listed in Table~\ref{tab:table2}).
At last, we put forward some conclusions corresponding to
Propositions 1-4.

{\it Proposition 5.} Let $E_a$ be an assisted bipartite entanglement measure. If it is polygamous for pure state,
then $E_a^\beta$ is polygamous for both pure and mixed states for any $0\leq\beta\leq1$.

{\it Proposition 6.} Let $E$ be a convex roof extended bipartite entanglement measure.
If $E_a^{\gamma}$ is polygamous for any tripartite system, then
$E_a^\beta$ is polygamous for any $n$-partite system $n\geq3$
and any $0\leq\beta\leq\gamma$. If $E$ is polygamous for all tripartite pure
states in $d\otimes d\otimes d'$ system, $d'=d^m$, $2\leq m\leq n-2$,
then it is polygamous for all $d^{\otimes n}$ states.

We now consider the convex roof extended of $E_a^\beta$. For any given entanglement measure $E$,
with the dual idea of $\tilde{E}^\alpha$ in mind, $\tilde{E}_a^\beta$
is defined similar as $\tilde{E}^\alpha$ by replacing the inf by sup for mixed states in Eq.~(\ref{crm})
and leaving the pure states invariant.
As one may expected, $\tilde{E}_a^\beta$ have the following properties.

{\it Proposition 7.} If $\tilde{E}_a^\beta$ is polygamous for pure states, then
it is also polygamous for mixed states.

{\it Proof.} Analogy to Proposition 3, we only need to check it for the tripartite case.
Let $\rho_{a\textbf{b}}$ be any given mixed state in $\mathcal{S}_{a\textbf{b}}$.
For any $\epsilon>0$, there exits an ensemble
$\{p_i,|\psi_i\ra\}$ such that
$\tilde{E}_a^\beta(\rho_{a|\textbf{b}})\leq\sum_ip_i\tilde{E}_a^\beta(\rho^{(i)}_{a|\textbf{b}})+\epsilon$,
where $\rho^{(i)}_{a\textbf{b}}=|\psi_i\ra\la\psi_i|$.
Then
$\tilde{E}_a^\beta(\rho_{a|\textbf{b}})\leq\sum_ip_i\tilde{E}_a^\beta(\rho^{(i)}_{a|\textbf{b}})+\epsilon
\leq\sum_ip_i[\tilde{E}_a^\beta(\rho^{(i)}_{a|b_1})+\tilde{E}_a^\beta(\rho^{(i)}_{a|b_2})]+\epsilon
\leq\tilde{E}_a^\beta(\rho_{a|b_1})+\tilde{E}_a^\beta(\rho_{a|b_2})+\epsilon$,
which complete the proof.
\hfill$\square$

\begin{table}
\caption{\label{tab:table2}The comparison of the polygamy power of several assisted entanglement.}
\begin{ruledtabular}
\begin{tabular}{cccc}
 $E_a$& $\beta(E_a)$ & system & reference \\ \colrule
$C_a$ & $\geq2$& $2^{\otimes 3}$ & \cite{Gour,Gour2005pra}\footnotemark[1]\\
$N_a$& $\geq 2$ & $2^{\otimes n}$ &\cite{Kim2009}\footnotemark[1]\\
 ${E_f}_a$& $\geq 1$ & any systems & \cite{Buscemi,Kim2012pra}\\
$\tau_a$& $\geq 1$ & $2^{\otimes n}$ &\cite{Lizongguo}\\
${T_q}_a$, $q\ge1$ &   $\geq 1$    & any system & \cite{Kim2010pra,Kim2016pra}\\
\end{tabular}
\end{ruledtabular}
\footnotetext[1]{For pure states.}
\end{table}

{\it Proposition 8.} If $E_a^2$ is polygamous, then
so is $\tilde{E}_a^2$.

{\it Proof.} It is sufficient to prove $E_a^2\leq \tilde{E}_a^2$
for mixed states.
Let $\rho_{ab}$ be any state in $\mathcal{S}_{ab}$.
For any $\epsilon>0$, there exists
$\{p_i,|\psi_i\rangle\}$ such that
$\rho_{ab}=\sum\limits_ip_i|\psi_i\rangle\langle\psi_i|$ and
$E_a(\rho_{ab})\leq\sum\limits_ip_iE(|\psi_i\rangle)+\epsilon$.
 Then we have
$\tilde{E}_a^2(\rho_{ab})\geq\sum\limits_ip_iE^2(|\psi_i\rangle)
=[\sum\limits_ip_i][\sum\limits_ip_iE^2(|\psi_i\rangle)]
\geq[\sum\limits_i\sqrt{p_i}\sqrt{p_i}E(|\psi_i\rangle)]^2
=[\sum\limits_ip_iE(|\psi_i\rangle)]^2
=(E_a(\rho_{ab})-\epsilon)^2$,
which yields $E^2\leq\tilde{E}^2$.\hfill$\square$

In conclusion, we depicted the monogamy relation for entanglement and
the polygamy relation for assisted entanglement in terms of the monogamy power and
the polygamy power respectively.
The monogamy power is the accurate critical value for the powers of entanglement
measure to be monogamous while the polygamy power is that of the assisted entanglement.
We showed the existence of theses two power indexes
(The former one is conditioned on the Conjecture. It is true at least for
the known systems, for example, the multiqubit system, and also true for $E_d$, $E_{sq}$
and $N_{cr}$
irrespective of the systems,
even though it may be false for other entanglement measures of higher dimension systems).
This improves the results in Ref.~\cite{Salini}
that based on the monotonically increasing functions of quantum correlations,
the general function $f(E(\rho_{ab_1}),E(\rho_{ab_2}))$ proposed in \cite{Lancien}
and the polynomial relation (for negativity) \cite{Allen},
which can not
lead to such an exact value.
Therefore, the monogamy and polygamy problems are reduced to find the critical values,
i.e., the monogamy power and the polygamy power.
We thus established, on general grounds, a new sketch of analyzing both the entanglement measure itself
and the distribution of entanglement.
Of course, our approach can also be applied to studying other bipartite quantum correlation measures,
such as quantum discord \cite{Ollivier}, quantum deficit \cite{Oppenheim},
measurement-induce nonlocality \cite{Luoshunlong}, quantum steering \cite{Wiseman}, etc.

\begin{acknowledgements}
This work was completed while Guo was visiting
the Institute of Quantum Science and
Technology of the University of Calgary
under the
support of China Scholarship Council.
Guo thanks Professor
C. Simon
and Professor G. Gour for their hospitality.
The author is supported by the Natural
Science Foundation of China under Grant No. 11301312.

\end{acknowledgements}



\end{document}